# The common revenue allocation based on modified Shapley value and DEA cross-efficiency

Xinyu Wang[a], Qianwei Zhang[a,*], Binwei Gui[b], Yingdi Zhao[a]

[a]School of Mathematics, Renmin University of China, Beijing, 100872 PR China

[b]Center for Law and Economics, China University of Political Science and Law, Beijing, 100088 PR China

**Abstract**

How to design a fair and reasonable allocation plan for the common revenue of the alliance is considered in this paper. We regard the common revenue to be allocated as an exogenous variable which will not participate the subsequent production process. The production organizations can cooperate with each other and form alliances. As the DEA cross-efficiency combines self- and peer-evaluation mechanisms, and the cooperative game allows fair negotiation among participants, we combine the cross-efficiency with the cooperative game theory and construct the modified Shapley value to reflect the contribution of the evaluated participant to the alliance. In addition, for each participant, both the optimistic and the pessimistic modified Shapley values are considered, and thus the upper and lower bounds of the allocation revenue are obtained, correspondingly. Numerical example is presented to illustrate the operation procedure. Finally, we apply the approach to an empirical application concerning a city commercial bank with 18 branches in China.

## 1. Introduction

Managers of large enterprises often face the problem of how to design a fair and reasonable scheme to allocate the common revenue or fixed costs among their branches. In order to be accepted by all participants, the allocation scheme should reflect both the overall fairness and the satisfaction of each participant. Because the total amount of the allocated revenue is fixed, the revenue allocated to one participant (i.e., decision making unit, DMU) will directly affect the revenue obtained by others. As a result, the participants in the alliance are usually with both cooperative and competitive relationships. Therefore, the DEA-game (data envelopment analysis) method is very applicable in allocation problems.

    DEA was first proposed by Charnes et al. [1] in 1978, and it has been extensively applied in performance evaluation and other management decision-making fields [2–4]. For example, Li et al. [5] considered the target setting and obtained the efficiencies of DMUs based on a common set of weights that are unchanged before and after the resource allocation, and received the resource allocation schemes. Nakabayashi and Tone [6] proposed a DEA game approach to solve the egoist's dilemma and their analytical framework can be used for fixed cost allocation and benefit distribution. Xiong et al. [7] considered the leader-follower relationship between two sub-units in a non-cooperative model to investigate the resource allocation problem in a bidirectional

interactive parallel system. Based on the principle of efficiency invariance, An et al. [8] proposed a fixed cost allocation method for systems with two-stage. They studied cooperative and non-cooperative scenarios based on the overall efficiency invariance and the partition efficiency invariance principles, respectively. Li et al. [9] adopted the nucleolus concept and proposed an approach for fixed cost allocation. Meng et al. [10] incorporated the perspectives of the coalition efficiency and the Shapley value, and gave an acceptable range of each coalition's allocated fixed cost. They obtained the final cost allocation results based on three principles: efficiency, monotonicity, and similarity.

Cross-efficiency method proposed by Sexton et al. [11] is often utilized in the design of allocation scheme. It combines the self-evaluation and the peer-evaluation by using reasonable weights. With the property of the peer-appraisal, the cross-efficiency is considered as a reasonable and appropriate mechanism for allocating the common revenue or the fixed cost. Using the cross-efficiency iterative algorithm, Du et al. [12] proposed a fixed cost allocation method with the goal of maximizing the cross-efficiency for all DMUs after assigning the cost. They proved that the approach is feasible and that the optimal cost allocation plan can render all DMUs efficient. For allocating revenues or fixed costs, Dai et al. [13] proposed a two-step incentive approach. Using a DEA game cross-efficiency method, they measured the efficiency scores of DMUs with non-cooperative relationship. Based on the performanceevaluation, they proposed nonlinear programme allocation incentive models and explored several properties. Considering the allocated cost as an additional input, Li et al. [14] proposed a DEA-game cross-efficiency approach and generated a unique and equitable allocation plan. In their method, each DMU is treated as a player and both the cooperative and competitive relationships of DMUs are considered. Taking the peer appraisal into account, they obtained the cross-efficiency of each DMU. Then, the Shapley value is calculated for each DMU and accordingly the optimal common weights are obtained to determine the fair distribution scheme which is acceptable by all DMUs. Their approach is "promising and attractive" in the issue of fixed cost allocation for large enterprises. Considering individual rationality and fairness, Chu et al. [15] proposed a new fixed cost allocation method based on cross-efficiency. For individual rationality, they developed two principles: a novel self-lowest principle and the efficient after fixed cost allocation. Furthermore, for overall fairness, the lower bound and cross-efficiency Pareto-optimality principles are proposed. Based on these principles, a multi-objective model is developed to obtain the final allocation results.

By surveying the previous literature, it is not difficult to find that in the issue of resource allocation, almost all of the existing literature considers the allocated resource as an additional input, and designs allocation schemes following the principle of efficiency invariance or efficiency maximization. However, in reality, the allocated resources are usually not included in the input or output of the subsequent production process of the DMU. For example, enterprises pay bonuses according to the employees' performance, and production organizations distribute profits in accordance with the operation performance of each branch and so on. These awards should be regarded as neither input nor output, and they should not be added to the subsequent performance evaluation. As a result, these bonuses and profits should be regarded as the exogenous variables. Yang and Zhang [16] gave the characteristic function with super-additivity property based on the out-oriented DEA efficiency, and proposed a modified Shapley value to allocate the resources in a fair way. Furthermore, they constructed the Gini coefficient to show that their resource allocation scheme is fairer than the other existing allocation schemes. In addition, when the input and (or) output are with fuzzy data, they proposed a method on the resource allocation scheme among DMUs [17]. However, because Yang and Zhang [16] adopted the traditional self-evaluation rather than the cross-efficiency to design the characteristic function, this treatment will have a certain impact on the fairness of the distribution.

In this paper we consider peer appraisal among DMUs and obtain characteristic functions based on DEA cross-efficiency. The modified Shapley value as well as its upper and lower bounds are designed to reflect the contribution of each DMU to the production alliance. Based on the modified Shapley value and its upper and lower bounds, we can obtain the range of the common revenue distribution for each DMU.

The rest of this paper is organized as follows. In Section 2, we review the DEA cross-efficiency method and introduce the game cross-efficiency model proposed by Yang et al. [18]. This model guarantees the uniqueness of the weights. In Section 3, we design a characteristic function with super-additivity and construct the modified Shapley value. Furthermore, both the optimistic and the pessimistic modified Shapley values are considered, and thus we can obtain the upper and lower bounds of the allocated income of each DMU, correspondingly. In Section 4, we apply the proposed approach to a numerical example to illustrate the specific procedure of our method. An empirical application of 18 commercial bank branches' activity is considered at the end of this section. Finally, concluding remarks are made in Section 5.

## 2. DEA cross-efficiency

Adopting the conventional nomenclature of DEA, we suppose that the production organization has $n$ DMUs with comparable and homogeneous characteristics, and each $DMU_j$ ($j = 1, 2, \ldots, n$) consumes $m$ inputs to produce $s$ outputs. The input and output vectors of $DMU_j$ are denoted as $\tilde{X}_j = (\tilde{x}_{1j}, \tilde{x}_{2j}, \ldots, \tilde{x}_{mj})$ and $\tilde{Y}_j = (\tilde{y}_{1j}, \tilde{y}_{2j}, \ldots, \tilde{y}_{sj})$, respectively. In order to take the size of each DMU into consideration, we normalize the input-output data and transfer the inputs and outputs to the following measures [19]:

$$x_{ij} = \frac{\tilde{x}_{ij}}{\sum_{j=1}^{n} \tilde{x}_{ij}}, \qquad y_{kj} = \frac{\tilde{y}_{kj}}{\sum_{j=1}^{n} \tilde{y}_{kj}}, \quad i = 1, 2, \ldots, m, \quad k = 1, 2, \ldots, s.$$

Based on the normalized data mentioned above, the efficiency value of the evaluated $DMU_d$ can be calculated by the following CCR model:

$$\max \frac{\sum_{r=1}^{s} u_r y_{rd}}{\sum_{i=1}^{m} v_i x_{id}} = \theta_d$$

$$\text{s.t.} \quad \frac{\sum_{r=1}^{s} u_r y_{rd}}{\sum_{i=1}^{m} v_i x_{id}} \leq 1, \ j = 1, 2, \ldots, n, \tag{1}$$

$$u_r \geq 0, \ v_i \geq 0, \ r = 1, 2, \ldots, s; \ i = 1, 2, \ldots, m.$$

For each evaluated $DMU_d$ ($d = 1, 2, \ldots, n$), we can obtain a set of optimal weights. Suppose that $u^{d*} = (u_1^{d*}, u_2^{d*}, \ldots, u_m^{d*})$ and $v^{d*} = (v_1^{d*}, v_2^{d*}, \ldots, v_s^{d*})$ are optima solutions of model (1) and the optimal value is $\theta_d^*$. It is obvious that these solutions are most preferred by $DMU_d$. Using the weights preferred by $DMU_d$, the cross-efficiency of any $DMU_j$ ($j = 1, 2, \ldots, n$) can be calculated by the following formula (2),

$$E_{d,j} = \frac{\sum_{r=1}^{s} u_r^{d*} y_{rj}}{\sum_{i=1}^{m} u_i^{d*} x_{ij}}, \quad j = 1, 2, \ldots, n. \tag{2}$$

$E_{d,j}$ is called $d$-cross-efficiency of $DMU_j$, and it represents the evaluation efficiency value of the $DMU_d$ for $DMU_j$. Since each $DMU_d$ ($d = 1, 2, \ldots, n$) has $n$ $d$-cross-efficiency scores, a cross-efficiency matrix can be obtained.

However, the weights are not unique, hence the cross-efficiency is not unique, which causes a great difficulty in using it to design a revenue allocation. Therefore, many scholars have paid attention to improving the cross-efficiency [20–23]. Here, we adopt the approach of Yang et al. [20]. They divided all DMUs into $H$ groups according to the hierarchical clustering method, where each DMU belongs to the corresponding type set $T_t$ ($t = 1, 2, \ldots, H$). DMUs belonging to the unified type set are allies, and DMUs that do not belong to the same type set are adversaries. The goal of $DMU_d$ is to maximize the efficiency of its allies and to minimize the efficiency of all of its adversaries under the premise that it reaches the optimal efficiency value $\theta_d^*$. The model is expressed as follows:

$$\min \sum_{j \neq d, j \in T_t} s_{dj} - \sum_{j \notin T_t} s_{dj}$$

s.t. $\sum_{r=1}^{s} u_r y_{rj} - \sum_{i=1}^{m} v_i x_{ij} + s_{dj} = 0, \ j = 1, 2, \cdots, n, j \neq d,$

$$\sum_{r=1}^{s} u_r y_{rj} - E_{d,j} \sum_{i=1}^{m} v_i x_{ij} = 0,$$

$j = 1, 2, \ldots, n, j \neq d,$  (3)

$$\sum_{r=1}^{s} u_r y_{rd} - \theta_d^* \sum_{i=1}^{m} v_i x_{id} = 0,$$

$s_{dj} \geq 0, 0 \leq E_{d,j} \leq 1, j = 1, 2, \ldots, n,$

$u_r \geq 0, v_i \geq 0, \ r = 1, 2, \ldots, s; \ i = 1, 2, \ldots, m.$

The optimal value of the nonlinear programming model described above is equal to that of the linear programming model below:

$$\min \sum_{j \neq d, j \in T_t} s_{dj} - \sum_{j \notin T_t} s_{dj}$$

s.t. $\sum_{r=1}^{s} u_r y_{rj} - \sum_{i=1}^{m} v_i x_{ij} + s_{dj} = 0, \ j = 1, 2, \ldots, n, j \neq d$

$$\sum_{r=1}^{s} u_r y_{rd} - \theta_d^* \sum_{i=1}^{m} v_i x_{id} = 0,$$  (4)

$s_{dj} \geq 0, 0 \leq E_{d,j} \leq 1, j = 1, 2, \ldots, n,$

$u_r \geq 0, v_i \geq 0, \ r = 1, 2, \ldots, s; \ i = 1, 2, \ldots, m.$

When the model (4) obtains the optimal solution $(\hat{u}_r, \hat{v}_i, \hat{s}_{dj})$, denoting

$$E_{d,j} = \frac{\sum_{r=1}^{s} \hat{u}_r y_{rj}}{\sum_{i=1}^{m} \hat{v}_i x_{ij}}, \quad d, j = 1, 2, \ldots, n,$$  (5)

then $E_{d,j}$ is the unique cross-efficiency of $DMU_j$ related to the $DMU_d$ when considering the cooperative and competitive relationships simultaneously.

## 3. Modified Shapley value

In this section, we combine the DEA cross-efficiency method and the cooperative game theory, and design the modified Shapley value based on the cross-efficiency to reflect the contribution of the DMU to the alliance.

### 3.1. Characteristic function

Suppose that a decision maker of an organization with $n$ branches (here they can be considered as DMUs) needs to distribute the common revenue fairly within the organization. The appropriate allocation manner means that the allocated proportion or quantity should be accepted by each participant. Therefore, the distribution of the income of each participant should be consistent with his contribution to the alliance. Shapley value just meets the needs. In the following, we first give the definition of cross-efficiency of the alliance.

**Definition 1.** For any participant $j$ which belongs to the coalition $S$, the upper and lower bounds on the cross-efficiency of $j$ are defined respectively as follows:

$$\overline{e}_{S,j}^{cross} = \max_{d \in S}\{E_{d,j}, d \neq j\} \tag{6}$$

and

$$\underline{e}_{S,j}^{cross} = \min_{d \in S}\{E_{d,j}, d \neq j\} \tag{7}$$

In particular, if there is only one participant $j$ in the coalition $S$, it is obvious that the equality $\overline{e}_{S,j}^{cross} = \underline{e}_{S,j}^{cross} = 1$ holds.

From Definition 1, we can see that the upper bound of the cross-efficiency of the $DMU_j$ is the maximum value of the evaluation efficiency of the other DMUs in the alliance on the $DMU_j$. Similarly, the lower bound of cross-efficiency is the minimum value of the evaluation efficiency values.

Following the above definition, we give specific representations of the characteristic function $v(S)$ as follows.

**Definition 2.** For any coalition $S$ which is the subset of the grand coalition $N$, the characteristic function is defined as

$$v(S) = \sum_{j \in S} \overline{e}_{S,j}^{cross} = \sum_{j=1}^{s} \overline{e}_{S,j}^{cross}, \tag{8}$$

where $s$ represents the number of participants in the coalition $S$.

Definition 2 means that the characteristic function $v(S)$ is the sum of the upper bounds on the cross-efficiency of each participant within the coalition $S$. For such a design, we can show that the characteristic function is super-additive.

**Theorem 1.** The characteristic function $v(S)$ of Definition 2 is super-additive, i.e., for any $S_1 \subseteq N$ and $S_2 \subseteq N$ with $S_1 \cap S_2 = \emptyset$, it holds

$$v(S_1 \cup S_2) \geq v(S_1) + v(S_2). \tag{9}$$

Proof. For any $S_1 \subseteq N$ and $S_2 \subseteq N$ with $S_1 \cap S_2 = \emptyset$, based on Definitions 1 and 2, we have

$$v(S_1) = \sum_{j \in S_1} \overline{e}_{S_1,j}^{cross} = \sum_{j=1}^{s_1} \overline{e}_{S_1,j}^{cross} = \sum_{j=1}^{s_1} \max_{d \in S_1}\{E_{d,j}, d \neq j\},$$

$$v(S_2) = \sum_{j \in S_2} \overline{e}_{S_2,j}^{cross} = \sum_{j=1}^{s_2} \overline{e}_{S_2,j}^{cross} = \sum_{j=1}^{s_2} \max_{d \in S_2}\{E_{d,j}, d \neq j\},$$

$$v(S_1 \cup S_2) = \sum_{j \in S_1 \cup S_2} \overline{e}_{S_1 \cup S_2,j}^{cross} = \sum_{j=1}^{s_1+s_2} \overline{e}_{S_1 \cup S_2,j}^{cross} = \sum_{j=1}^{s_1+s_2} \max_{d \in S_1 \cup S_2}\{E_{d,j}, d \neq j\}.$$

For any $j \in S_1 \cup S_2$, we have

$$\sum_{j=1}^{s_1} \max_{d \in S_1}\{E_{d,j}, d \neq j\} \leq \max_{d \in S_1 \cup S_2}\{E_{d,j}, d \neq j\}$$

and

$$\sum_{j=1}^{s_2} \max_{d \in S_2}\{E_{d,j}, d \neq j\} \leq \max_{d \in S_1 \cup S_2}\{E_{d,j}, d \neq j\}.$$

So we can see that the following equalities and inequalities hold:

$$v(S_1 \cup S_2) = \sum_{j=1}^{s_1+s_2} \max_{d \in S_1 \cup S_2} \{E_{d,j}, d \neq j\}$$

$$= \sum_{j=1}^{s_1} \max_{d \in S_1 \cup S_2} \{E_{d,j}, d \neq j\} + \sum_{j=1}^{s_2} \max_{d \in S_1 \cup S_2} \{E_{d,j}, d \neq j\}$$

$$\geq \sum_{j=1}^{s_1} \max_{d \in S_1} \{E_{d,j}, d \neq j\} + \sum_{j=1}^{s_2} \max_{d \in S_2} \{E_{d,j}, d \neq j\}$$

$$= v(S_1) + v(S_2).$$

Thus, the super-additivity of the characteristic function $v(S)$ is proved.

### 3.2. Modified Shapley value

According to the cooperative game theory, we know that the traditional Shapley value based on the characteristic function $v(S)$ is defined by the formula [24].

$$\varphi_i(v) = \sum_{S \subseteq N} \frac{s!(n-s-1)!}{n!} \cdot [v(S \cup \{i\}) - v(S)], \quad i \notin S, \tag{10}$$

where for any $i \in N$, $s$ and $n$ are the numbers of participants in $S$ and $N$, respectively.

The value of the above formula means the contribution of the participant $i$ in the cooperative game. For an $n$-person game, the Shapley value is expressed by an $n$-dimensional vector. Each component of the vector indicates the extent the corresponding participant contributes to the grand coalition $N$. For the common revenue allocation problem, we can design the modified Shapley value according to the meaning of the Shapely value and the peer evaluation character of DEA cross-efficiency.

**Definition 3.** For a grand coalition $N$ with $n$ participants, the modified Shapley value is an $n$-dimensional vector $\Phi(v) = [\varphi_1(v), \varphi_2(v), \ldots, \varphi_n(v)]$, where the component is

$$\varphi_i(v) = \sum_{S \subseteq N} \frac{s!(n-s-1)!}{n!} \cdot \frac{1 + \overline{e}^{cross}_{S \cup \{i\},i} - \overline{e}^{cross}_{i,i}}{s + \sum_{j \in S} \overline{e}^{cross}_{S \cup \{i\},j} - \sum_{j \in S} \overline{e}^{cross}_{S,j}}, \quad i \notin S, i = 1,2,\ldots,n. \tag{11}$$

It should be noted that in the above equation (11), the $[v(S \cup \{i\}) - v(S)]$ traditionally given in the Shapley value is transformed into a fractional expression, where the denominator represents the influence of the $DMU_i$ on the original cross-efficiencies of DMUs after it joins the alliance $S$; the numerator represents the influence of the $DMU_i$ on the cross-efficiency value of its own before and after it joins the alliance. Since $\overline{e}^{cross}_{S \cup \{i\},i} - \overline{e}^{cross}_{i,i}$ ($i \notin S, i = 1,2,\cdots,n$) must be located in the interval $(-1, 0]$ and $\overline{e}^{cross}_{S \cup \{i\},j} - \overline{e}^{cross}_{S,j}$ may be located within it, where in the equation (11) the 1 and $s$ are added to the numerator and the denominator, respectively, to ensure that both the numerator and the denominator are positive. Due to that $\overline{e}^{cross}_{i,i} = 1$, the formula (11) can be written in an equivalent form as follows,

$$\varphi_i(v) = \sum_{S \subseteq N} \frac{s!(n-s-1)!}{n!} \cdot \frac{\overline{e}^{cross}_{S \cup \{i\},i}}{s + \sum_{j \in S} \overline{e}^{cross}_{S \cup \{i\},j} - \sum_{j \in S} \overline{e}^{cross}_{S,j}}, \quad i \notin S, i = 1,2,\ldots,n. \tag{12}$$

For $DMU_i$, if the numerator is larger, then the cross-efficiency value of $DMU_i$ obtained after it joins the alliance $S$ is larger, which means that it is highly evaluated by the original members of the alliance $S$. At the same time, if $DMU_i$ joining $S$ can reduce the overall cross-efficiency values of other original DMUs in the alliance, i.e., the denominator decreases, then it shows that the $DMU_i$ has a large role and a high contribution to the alliance $S$, and thus the corresponding Shapley value is large. These are reflected by the equation (12).

Following the above design ideas of the modified Shapley values and the upper and lower bounds of the cross-efficiency, we can further give the definition of the upper and lower bounds of the modified Shapley values.

**Definition 4.** For any participant $i$ which belongs to the coalition $S$, the upper and lower bounds on the modified Shapley value of $i$ are defined respectively as follows,

$$\overline{\varphi}_i(v) = \sum_{S \subseteq N} \frac{s!(n-s-1)!}{n!} \cdot \frac{1 + \overline{e}_{S\cup\{i\},i}^{cross} - \underline{e}_{i,i}^{cross}}{s + \sum_{j \in S} \underline{e}_{S\cup\{i\},j}^{cross} - \sum_{j \in S} \overline{e}_{S,j}^{cross}}, \quad i \notin S, i = 1, 2, \ldots, n. \quad (13)$$

and

$$\underline{\varphi}_i(v) = \sum_{S \subseteq N} \frac{s!(n-s-1)!}{n!} \cdot \frac{1 + \underline{e}_{S\cup\{i\},i}^{cross} - \overline{e}_{i,i}^{cross}}{s + \sum_{j \in S} \overline{e}_{S\cup\{i\},j}^{cross} - \sum_{j \in S} \underline{e}_{S,j}^{cross}}, \quad i \notin S, i = 1, 2, \ldots, n. \quad (14)$$

The upper and lower bounds of the modified Shapley value represent the maximum and minimum values of the degree to which the $DMU_i$ contributes to the coalition $S$, respectively. The numerator $\overline{e}_{S\cup\{i\},i}^{cross} - \underline{e}_{i,i}^{cross}$ in formula (13) indicates the difference between the maximum cross-efficiency and the minimum value of the self-evaluated efficiency of the $DMU_i$ before and after it joins the alliance. The larger this value is, the higher recognition of the $DMU_i$ by the members in the alliance $S$. The $\sum_{j \in S} \overline{e}_{S\cup\{i\},j}^{cross} - \sum_{j \in S} \underline{e}_{S,j}^{cross}$ in the denominator is the difference between the sum of the lower bounds of the cross-efficiency values and the sum of the upper bounds of the cross-efficiency values of the original DMUs in the alliance $S$ before and after $DMU_i$ joins $S$. The smaller the value of $\sum_{j \in S} \underline{e}_{S\cup\{i\},j}^{cross}$ is, the higher the status of the $DMU_i$ in the alliance $S$, because its participation reduces the peer evaluation of the other DMUs. Similar to the numerator and denominator in formula (13), the design of the numerator and denominator in formula (14) describes the minimum contribution of the $DMU_i$ to the alliance $S$.

Since $\overline{e}_{i,i}^{cross} = \underline{e}_{i,i}^{cross} = 1$ holds, the formulae (13) and (14) can be written in the following equivalent forms accordingly:

$$\overline{\varphi}_i(v) = \sum_{S \subseteq N} \frac{s!(n-s-1)!}{n!} \cdot \frac{\overline{e}_{S\cup\{i\},i}^{cross}}{s + \sum_{j \in S} \underline{e}_{S\cup\{i\},j}^{cross} - \sum_{j \in S} \overline{e}_{S,j}^{cross}}, \quad i \notin S, i = 1, 2, \ldots, n, \quad (15)$$

and

$$\underline{\varphi}_i(v) = \sum_{S \subseteq N} \frac{s!(n-s-1)!}{n!} \cdot \frac{\underline{e}_{S\cup\{i\},i}^{cross}}{s + \sum_{j \in S} \overline{e}_{S\cup\{i\},j}^{cross} - \sum_{j \in S} \underline{e}_{S,j}^{cross}}, \quad i \notin S, i = 1, 2, \ldots, n. \quad (16)$$

According to Definitions 3 and 4, we have the following conclusion.

**Theorem 2.** For the modified Shapley values of Definitions 3 and 4, the following relationships are established,

$$\underline{\varphi}_i(v) \leq \varphi_i(v) \leq \overline{\varphi}_i(v) \quad (17)$$

**Proof.** For any $i \notin S, i = 1, 2, \ldots, n$, we have

$$\underline{e}^{cross}_{S\cup\{i\},i} \leq \overline{e}^{cross}_{S\cup\{i\},i}, \quad \sum_{j\in S}\underline{e}^{cross}_{S,j} \leq \sum_{j\in S}\overline{e}^{cross}_{S,j}$$

and

$$\sum_{j\in S}\underline{e}^{cross}_{S\cup\{i\},j} \leq \sum_{j\in S}\overline{e}^{cross}_{S\cup\{i\},j}.$$

So the following inequalities hold

$$\frac{\underline{e}^{cross}_{S\cup\{i\},i}}{s+\sum_{j\in S}\overline{e}^{cross}_{S\cup\{i\},j}-\sum_{j\in S}\underline{e}^{cross}_{S,j}} \leq \frac{\overline{e}^{cross}_{S\cup\{i\},i}}{s+\sum_{j\in S}\overline{e}^{cross}_{S\cup\{i\},j}-\sum_{j\in S}\overline{e}^{cross}_{S,j}} \leq \frac{\overline{e}^{cross}_{S\cup\{i\},i}}{s+\sum_{j\in S}\underline{e}^{cross}_{S\cup\{i\},j}-\sum_{j\in S}\overline{e}^{cross}_{S,j}}.$$

As a result, there is $\underline{\varphi}_i(v) \leq \varphi_i(v) \leq \overline{\varphi}_i(v), i \notin S, i = 1,2,\ldots,n$.

### 3.3. Allocation scheme

For a production organization with $n$ DMUs, we can allocate the exogenous common revenue $R$ according to the modified Shapley value. It should be noted that the modified Shapley value of $DMU_i$ reflects the degree of its contribution to the organization that is not the final allocation result. Therefore, the proportion of revenue allocation obtained for the $DMU_i$ is determined as

$$\frac{\varphi_i(v)}{\sum_{i=1}^n \varphi_i(v)}, i = 1,2,\ldots,n. \tag{18}$$

Then the allocation scheme of the exogenous common revenue $R$ among the $n$ DMUs is designed by the following $n$-dimensional vector

$$T = \left(\frac{R\varphi_1(v)}{\sum_{i=1}^n \varphi_i(v)}, \frac{R\varphi_2(v)}{\sum_{i=1}^n \varphi_i(v)}, \ldots, \frac{R\varphi_n(v)}{\sum_{i=1}^n \varphi_i(v)}\right). \tag{19}$$

Furthermore, according to the upper and lower bounds of the modified Shapley values, we can further obtain the optimistic and pessimistic schemes of the common revenue allocation for each DMU as follows, respectively. The optimistic revenue allocation for $DMU_i$ $i = 1,2,\cdots,n)$ is

$$\frac{R\overline{\varphi}_i(v)}{\overline{\varphi}_i(v) + \sum_{k\in N, k\neq i}\underline{\varphi}_k(v)}, i = 1,2,\ldots,n, \tag{20}$$

and the pessimistic scheme for $DMU_i$ is

$$\frac{R\underline{\varphi}_i(v)}{\underline{\varphi}_i(v) + \sum_{k\in N, k\neq i}\overline{\varphi}_k(v)}, i = 1,2,\ldots,n. \tag{21}$$

## 4. Illustrations

In this section, we show the computational procedure of the revenue allocation according to the modified Shapley value based on the cross-efficiency with a numerical example. Then we apply this approach to an empirical revenue allocation concerning 18 commercial bank branches.

### 4.1. A numerical example

Suppose that a production organization is composed of five DMUs, and each DMU has three inputs and two outputs. The data set is shown in Table 1.

**Table 1**
The simple data set

|         | $DMU_1$ | $DMU_2$ | $DMU_3$ | $DMU_4$ | $DMU_5$ |
|---------|---------|---------|---------|---------|---------|
| Input 1 | 23 | 60 | 44 | 40 | 70 |
| Input 2 | 24 | 40 | 69 | 30 | 90 |
| Input 3 | 122 | 150 | 120 | 70 | 175 |
| Output 1 | 38 | 45 | 76 | 52 | 63 |
| Output 2 | 60 | 85 | 43 | 42 | 74 |

Based on the model (4) we can obtain the cross-efficiency matrix for the five DMUs as follows in Table 2.

**Table 2**

The cross-efficiency matrix

| Evaluator $DMU_j$ | Targeted DMU | | | | |
|---|---|---|---|---|---|
| | 1 | 2 | 3 | 4 | 5 |
| 1 | 1.00 | 0.46 | 0.25 | 0.40 | 0.33 |
| 2 | 0.89 | 1.00 | 0.40 | 0.81 | 0.51 |
| 3 | 0.58 | 0.43 | 1.00 | 0.75 | 0.52 |
| 4 | 0.42 | 0.40 | 0.56 | 1.00 | 0.40 |
| 5 | 1.00 | 1.00 | 0.69 | 1.00 | 1.00 |

The number of coalitions that are formed by these five DMUs is $2^5 - 1 = 31$. Take the coalitions $S = \{1,2,3\}$ for example. Judging from the formulae (6) and (7), we can get the upper and lower bounds on the cross-efficiency of each DMU as follows:

$$\overline{e}_{S,1}^{cross} = max\{E_{2,1}, E_{3,1}\} = max\{0.89, 0.58\} = 0.89, \quad \underline{e}_{S,1}^{cross} = min\{E_{2,1}, E_{3,1}\} = 0.58;$$

$$\overline{e}_{S,2}^{cross} = max\{E_{1,2}, E_{3,2}\} = max\{0.46, 0.43\} = 0.46, \quad \underline{e}_{S,2}^{cross} = min\{E_{1,2}, E_{3,2}\} = 0.43;$$

$$\overline{e}_{S,3}^{cross} = max\{E_{1,3}, E_{2,3}\} = max\{0.25, 0.40\} = 0.40, \quad \underline{e}_{S,1}^{cross} = min\{E_{1,3}, E_{2,3}\} = 0.25.$$

Furthermore, according to the equations (12), (15) and (16), we obtain the modified Shapley value as well as its upper and lower bounds for each DMU which are shown in Table 3.

**Table 3**

The modified Shapley values

|  | $DMU_1$ | $DMU_2$ | $DMU_3$ | $DMU_4$ | $DMU_5$ |
|---|---|---|---|---|---|
| $\overline{\varphi}_i(v)$ | 0.71 | 0.61 | 0.61 | 0.69 | 0.65 |
| $\varphi_i(v)$ | 0.31 | 0.19 | 0.15 | 0.27 | 0.09 |
| $\underline{\varphi}_i(v)$ | 0.22 | 0.11 | 0.09 | 0.18 | 0.06 |

If there are 10,000 units of revenue needed to be allocated fairly among the five DMUs, then according to the numerical results in Table 3, the formulae (18), (20) and (21) can provide the distribution results in Table 4 below.

**Table 4**

The results of revenue allocation

|  | $DMU_1$ | $DMU_2$ | $DMU_3$ | $DMU_4$ | $DMU_5$ |
|---|---|---|---|---|---|
| Upper bound | 3453.94 | 2182.48 | 1807.82 | 3027.706 | 1117.30 |
| Allocation scheme | 3065.85 | 1858.07 | 1510.14 | 2685.508 | 907.43 |

| Lower bound | 2716.22 | 1486.95 | 1207.85 | 2316.89 | 794.28 |

As we can see from Table 3, the $DMU_1$ has the largest modified Shapley value, i.e., the greatest contribution to the organization, and the $DMU_5$ is with the smallest modified Shapley value, so it has the least contribution to the league. In Table 4, it is not difficult to find that the revenue allocation results match the contribution degree of each DMU to the alliance, so $DMU_1$ gets the most revenue, and $DMU_5$ gets the least. The assigned result and its upper and lower bounds of each DMU are visually reflected by the histogram in Figure 1.

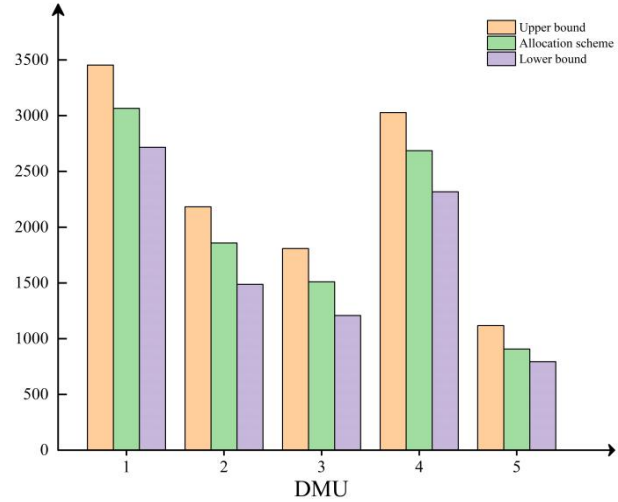

**Figure 1** Revenue allocation scheme

### 4.2. The empirical study

In this part, we apply the proposed method to the resource allocation of a city commercial bank with 18 branches in Sichuan Province of China. The input-output classification of these branches as well as the related data are referred to the references [25-27]. Unlike the existing researches, we distribute the exogenous resources (such as common revenue) fairly among the branches, based on the extent to which each of the 18 branches contributes to the city commercial bank. We select three inputs and three outputs similar to Li et al. [25] and these are summarized in Table 5.

**Table 5**
Input-output variables

|  | Variable | Interpretation | Unit |
|---|---|---|---|
| Input | $x_1$(staffs) | human resource | Person |
|  | $x_2$(fixed assets) | the asset value of physical capital | 10 thousand CNY |
|  | $x_3$(operating costs) | operating expenses other than employee expenses | 10 thousand CNY |
| Output | $y_1$(deposits) | includes current deposits and time deposits | 10 thousand CNY |
|  | $y_2$(loans) | loans given by the bank | 10 thousand CNY |
|  | $y_3$(revenue income) | Interest income and non-interest income | 10 thousand CNY |

The specific input-output data of these 18 commercial bank branches are derived from reference [25] and are given in Table 6.

**Table 6**

The input-output data of Li et al. [25]

| $DMU_j$ | $x_1$ | $x_2$ | $x_3$ | $y_1$ | $y_2$ | $y_3$ |
|---|---|---|---|---|---|---|
| 1 | 62 | 1822 | 1361 | 140117 | 130288 | 5260 |
| 2 | 80 | 1833 | 1565 | 213774 | 145761 | 10773 |
| 3 | 129 | 3595 | 1378 | 194084 | 130556 | 8006 |
| 4 | 62 | 1978 | 333 | 87876 | 49454 | 4479 |
| 5 | 89 | 2138 | 549 | 107091 | 60872 | 5897 |
| 6 | 84 | 1910 | 704 | 97472 | 94310 | 3849 |
| 7 | 36 | 1234 | 840 | 114001 | 80019 | 5292 |
| 8 | 172 | 4348 | 959 | 366423 | 306926 | 12479 |
| 9 | 62 | 879 | 1253 | 107393 | 86485 | 5132 |
| 10 | 53 | 2566 | 483 | 69691 | 43907 | 3869 |
| 11 | 92 | 1348 | 419 | 148458 | 87193 | 7234 |
| 12 | 39 | 1229 | 513 | 83752 | 40046 | 3984 |
| 13 | 144 | 4640 | 1323 | 223539 | 211466 | 10655 |
| 14 | 47 | 2248 | 670 | 70555 | 65110 | 2205 |
| 15 | 39 | 1571 | 362 | 99143 | 66736 | 5271 |
| 16 | 56 | 1635 | 669 | 112513 | 79366 | 5202 |
| 17 | 34 | 939 | 867 | 87660 | 56157 | 3000 |
| 18 | 58 | 1807 | 419 | 88334 | 67160 | 4171 |

The city commercial bank is regarded as a production organization, and each branch is considered as a homogeneous and independent DMU. Under the assumption that the branches can conduct mutual peer evaluation, the unique cross-efficiency matrix of these 18 branches can be obtained according to the model (4) and the formula (5), which is given as follows in Table 7.

**Table 7**

Cross-efficiency matrix of 18 commercial bank branches

| Evaluator $DMU_j$ | Targeted DMU | | | | | | | | | | | | | | | | | |
|---|---|---|---|---|---|---|---|---|---|---|---|---|---|---|---|---|---|---|
|  | 1 | 2 | 3 | 4 | 5 | 6 | 7 | 8 | 9 | 10 | 11 | 12 | 13 | 14 | 15 | 16 | 17 | 18 |
| 1 | 1.00 | 0.94 | 0.50 | 0.36 | 0.35 | 0.54 | 0.96 | 0.89 | 0.79 | 0.28 | 0.55 | 0.47 | 0.66 | 0.48 | 0.67 | 0.68 | 0.78 | 0.53 |
| 2 | 0.49 | 1.00 | 0.38 | 0.35 | 0.36 | 0.34 | 0.73 | 0.49 | 0.64 | 0.24 | 0.62 | 0.46 | 0.39 | 0.17 | 0.56 | 0.54 | 0.54 | 0.39 |
| 3 | 0.74 | 1.00 | 1.00 | 0.63 | 0.57 | 0.53 | 1.00 | 1.00 | 0.69 | 0.49 | 0.86 | 0.83 | 0.65 | 0.47 | 1.00 | 0.81 | 0.80 | 0.67 |
| 4 | 0.30 | 0.52 | 0.41 | 1.00 | 0.68 | 0.37 | 0.48 | 0.80 | 0.31 | 0.55 | 1.00 | 0.56 | 0.55 | 0.24 | 1.00 | 0.56 | 0.27 | 0.65 |
| 5 | 0.30 | 0.52 | 0.41 | 0.82 | 1.00 | 0.37 | 0.48 | 0.80 | 0.31 | 0.55 | 1.00 | 0.56 | 0.55 | 0.24 | 1.00 | 0.56 | 0.27 | 0.65 |
| 6 | 0.86 | 0.93 | 0.49 | 0.36 | 0.40 | 1.00 | 0.80 | 1.00 | 1.00 | 0.25 | 0.89 | 0.43 | 0.63 | 0.40 | 0.60 | 0.65 | 0.69 | 0.52 |
| 7 | 0.58 | 0.82 | 0.42 | 0.36 | 0.31 | 0.31 | 1.00 | 0.50 | 0.55 | 0.31 | 0.43 | 0.46 | 0.49 | 0.29 | 0.71 | 0.63 | 0.59 | 0.48 |
| 8 | 0.27 | 0.29 | 0.30 | 0.36 | 0.35 | 0.36 | 0.30 | 1.00 | 0.22 | 0.24 | 0.54 | 0.24 | 0.44 | 0.27 | 0.58 | 0.37 | 0.20 | 0.50 |
| 9 | 0.50 | 0.81 | 0.37 | 0.25 | 0.29 | 0.35 | 0.66 | 0.50 | 1.00 | 0.17 | 0.66 | 0.33 | 0.39 | 0.17 | 0.43 | 0.49 | 0.55 | 0.38 |
| 10 | 0.30 | 0.52 | 0.41 | 0.82 | 0.51 | 0.37 | 0.48 | 0.80 | 0.31 | 1.00 | 1.00 | 0.56 | 0.55 | 0.24 | 1.00 | 0.56 | 0.27 | 0.65 |
| 11 | 0.22 | 0.39 | 0.34 | 0.41 | 0.45 | 0.32 | 0.36 | 0.54 | 0.24 | 0.25 | 1.00 | 0.43 | 0.43 | 0.18 | 0.58 | 0.45 | 0.20 | 0.44 |
| 12 | 0.74 | 1.00 | 0.62 | 0.65 | 0.60 | 0.53 | 1.00 | 1.00 | 0.69 | 0.49 | 0.93 | 1.00 | 0.65 | 0.47 | 1.00 | 0.81 | 0.80 | 0.67 |
| 13 | 0.89 | 0.89 | 0.55 | 0.52 | 0.44 | 0.60 | 1.00 | 1.00 | 0.64 | 0.50 | 0.61 | 0.59 | 1.00 | 0.65 | 1.00 | 0.76 | 0.69 | 0.67 |
| 14 | 0.96 | 0.86 | 0.53 | 0.45 | 0.38 | 0.61 | 1.00 | 1.00 | 0.65 | 0.44 | 0.54 | 0.52 | 0.78 | 1.00 | 0.91 | 0.73 | 0.73 | 0.63 |
| 15 | 0.30 | 0.52 | 0.41 | 0.52 | 0.44 | 0.34 | 0.48 | 0.55 | 0.31 | 0.47 | 0.60 | 0.53 | 0.55 | 0.24 | 1.00 | 0.56 | 0.27 | 0.54 |
| 16 | 0.74 | 1.00 | 0.62 | 0.65 | 0.60 | 0.53 | 0.97 | 1.00 | 0.69 | 0.49 | 0.93 | 0.83 | 0.65 | 0.47 | 1.00 | 1.00 | 0.77 | 0.67 |
| 17 | 0.77 | 1.00 | 0.52 | 0.46 | 0.44 | 0.44 | 1.00 | 0.77 | 0.76 | 0.35 | 0.70 | 0.70 | 0.50 | 0.40 | 0.74 | 0.68 | 1.00 | 0.50 |
| 18 | 0.74 | 1.00 | 0.58 | 0.63 | 0.63 | 0.60 | 0.93 | 1.00 | 0.74 | 0.47 | 1.00 | 0.71 | 0.63 | 0.43 | 1.00 | 0.80 | 0.63 | 1.00 |

Based on these cross-efficiency values, we can get the corresponding modified Shapley value as well as its upper and lower bounds for each branch by calculating the formulae (12), (15) and (16). These values are shown in the following Table 8.

**Table 8**

Modified Shapley values for each branch

|  | $DMU_1$ | $DMU_2$ | $DMU_3$ | $DMU_4$ | $DMU_5$ | $DMU_6$ | $DMU_7$ | $DMU_8$ | $DMU_9$ |
|---|---|---|---|---|---|---|---|---|---|
| $\overline{\varphi}_i(v)$ | 0.24 | 0.32 | 0.16 | 0.22 | 0.19 | 0.17 | 0.30 | 0.37 | 0.27 |
| $\varphi_i(v)$ | 0.17 | 0.24 | 0.11 | 0.15 | 0.13 | 0.12 | 0.22 | 0.27 | 0.13 |
| $\underline{\varphi}_i(v)$ | 0.08 | 0.14 | 0.06 | 0.08 | 0.08 | 0.07 | 0.13 | 0.19 | 0.11 |
|  | $DMU_{10}$ | $DMU_{11}$ | $DMU_{12}$ | $DMU_{13}$ | $DMU_{14}$ | $DMU_{15}$ | $DMU_{16}$ | $DMU_{17}$ | $DMU_{18}$ |
| $\overline{\varphi}_i(v)$ | 0.16 | 0.34 | 0.20 | 0.19 | 0.14 | 0.32 | 0.21 | 0.21 | 0.19 |
| $\varphi_i(v)$ | 0.11 | 0.25 | 0.14 | 0.14 | 0.10 | 0.24 | 0.15 | 0.15 | 0.13 |
| $\underline{\varphi}_i(v)$ | 0.06 | 0.17 | 0.07 | 0.08 | 0.05 | 0.15 | 0.09 | 0.07 | 0.08 |

To facilitate the comparison with the reference [25], we assume that the revenue is $R = 2900$ units (1 unit = 10 thousand CNY) and it will be distributed fairly among the 18 branches. According to the degree of contribution of each branch to the city commercial bank, the distribution of these 2900 units as well as the allocation results in the optimistic and pessimistic situations are obtained by the formulae (18), (20) and (21), respectively. The allocation results of the branches are presented in Table 9.

**Table 9**

The results of the revenue allocation

|  | $DMU_1$ | $DMU_2$ | $DMU_3$ | $DMU_4$ | $DMU_5$ | $DMU_6$ | $DMU_7$ | $DMU_8$ | $DMU_9$ |
|---|---|---|---|---|---|---|---|---|---|
| Upper bound | 356.98 | 477.34 | 250.80 | 329.92 | 289.51 | 261.64 | 452.38 | 552.19 | 407.40 |
| Allocation scheme | 161.01 | 227.89 | 106.89 | 147.21 | 127.47 | 114.29 | 213.01 | 262.81 | 187.30 |
| Lower bound | 57.41 | 103.37 | 44.86 | 60.32 | 55.84 | 50.35 | 91.40 | 134.92 | 77.88 |
|  | $DMU_{10}$ | $DMU_{11}$ | $DMU_{12}$ | $DMU_{13}$ | $DMU_{14}$ | $DMU_{15}$ | $DMU_{16}$ | $DMU_{17}$ | $DMU_{18}$ |
| Upper bound | 246.09 | 515.18 | 307.75 | 301.04 | 220.83 | 482.42 | 328.05 | 320.82 | 289.00 |
| Allocation scheme | 107.98 | 244.64 | 133.98 | 130.75 | 93.02 | 228.66 | 143.71 | 143.67 | 125.72 |
| Lower bound | 44.99 | 121.17 | 50.70 | 57.27 | 31.93 | 108.93 | 61.42 | 51.79 | 56.47 |

For intuition, the revenue allocation results for the 18 branches together with the corresponding cost assignment results in Li et al. [25] are presented in Figure 2.

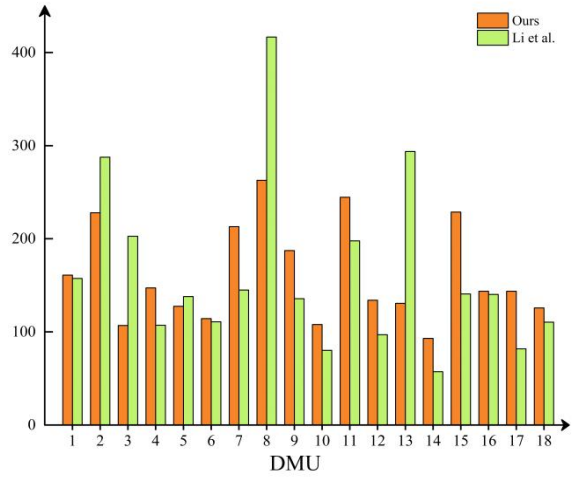

**Figure 2** Revenue and cost allocation results

According to Figure 2, it is easy to see that, for the 18 branches, our revenue allocation results are relatively similar to the cost-sharing results of Li et al. [25]. Regardless of revenue allocation or cost distribution, the assignment value of the $DMU_8$ is the largest, and that of the $DMU_{14}$ is the smallest. It is known from Tables 7 and 8 that although the cross-efficiency value of the $DMU_8$ is not as high as that of the $DMU_{15}$, the modified Shapley value of the $DMU_8$ is the highest among the 18 branches. Moreover, it is not difficult to find that the three outputs of the $DMU_8$, namely deposits, loans and revenue income, are the highest among the 18 branches. This phenomenon fully shows that the revenue allocation scheme designed by our proposed method, not only presents the size of cross-efficiency of the DMU, but also reflects the contribution of the DMU to the system by the modified Shapley value based on the cross-efficiency, and then the revenue is allocated to each DMU according to the degree of its contribution.

From the input-output data in Table 6, we can see that the input vector of the $DMU_{14}$ is greater than that of the $DMU_{15}$, while the output vector of it is less than that of the $DMU_{15}$. By the columns 14 and 15 of the cross-efficiency matrix which are expressed in Table 7, it is known that the peer evaluation cross-efficiency values of the $DMU_{14}$ are less than the corresponding cross-efficiency values of the $DMU_{15}$. Moreover, the total cross-efficiency value of the $DMU_{14}$ is the smallest one of the 18 branches, and the corresponding modified Shapley value of 0.10 is also the smallest in these 18 DMUs. As a result, using our modified Shapley value method, the allocation result of the $DMU_{14}$ is 0.9302 million yuan and the allocation result of $DMU_{15}$ is 2.2866 million yuan.

For the $DMU_7$ and the $DMU_{10}$, except the third type of inputs $x_3$ (operating expenses other than employee expenses), the inputs $x_1$, $x_2$ and $x_3$ of $DMU_7$ are less than the corresponding inputs of the $DMU_{10}$, while the outputs $y_1$ and $y_2$ of the $DMU_7$ are all greater than the corresponding outputs of the $DMU_{10}$. Therefore, it can be seen from Table 7 that the total cross-efficiency of the $DMU_7$ must be greater than that of the $DMU_{10}$. According to our distribution scheme, the income of the $DMU_7$ is 2.1301 million yuan, and that of the $DMU_{10}$ is only 1.0798 million yuan.

Furthermore, comparing the numerical of the $DMU_3$ and the $DMU_4$, and that of the $DMU_{11}$ and the $DMU_{13}$, we find that in the case of revenue allocation, the result of the $DMU_3$ is less than that of the $DMU_4$, while the cost allocation of the $DMU_3$ is higher than that of the

$DMU_4$ [25]. For the $DMU_{11}$ and the $DMU_{13}$, the revenue allocation result is also opposite to the cost allocation result. However, regardless of the cross-efficiency values in our Table 7 or those in Table 13 [25], they all show that the cross-efficiency of the $DMU_3$ ($DMU_{13}$) is less than that of the $DMU_4$ ($DMU_{11}$). Meanwhile, from the perspective of the modified Shapley value, our results are the modified Shapley values of 0.11 (0.14) and 0.15 (0.25) for the $DMU_3$ ($DMU_{13}$) is less than that of the $DMU_4$ ($DMU_{11}$), respectively. Therefore, the reason for this phenomenon lies in different view and method of research.

Moreover, another important feature of our allocation scheme is that the upper and lower allocation bounds are given for each DMU, which has a particularly important role in practical applications. Different from the models in the ideal case, in many realistic situations, there are many variable and unpredictable factors, which requires that the allocation scheme should have some flexibility in dealing with a variety of different situations, and this strengthens the advantage of our proposed method.

We give the upper and lower bounds on the distribution values for these 18 branches, and present the distribution results together with these of [25] in Figure 3. From Figure 3, we can see that the allocation amount of each DMU of [25] is between the upper and lower bounds, and the trend of the four-line charts is roughly consistent, which shows that the upper and lower bounds proposed by us are feasible for this revenue allocation scheme. By further observation we can find that the reference [25] shows that the cost allocation of the $DMU_{13}$ is 2.937 million yuan, and the allocation of our upper bound is 3.0104 million yuan; moreover, the allocation amount of the $DMU_{14}$ is 572.4 thousand yuan, and the lower allocation bound that we give is 319.3 thousand yuan. The comparative analysis shows that the upper and lower bounds proposed by us are in a reasonable range, and there is no occurrence of too high or too low situation. Therefore, the distribution of the revenue of the city commercial bank among 18 branches can be based on the amount of allocation of our suggested method, and the allocation results can be flexibly adjusted according to the reality and other factors on the premise of all the distribution results between the upper and lower bounds mentioned in Table 9.

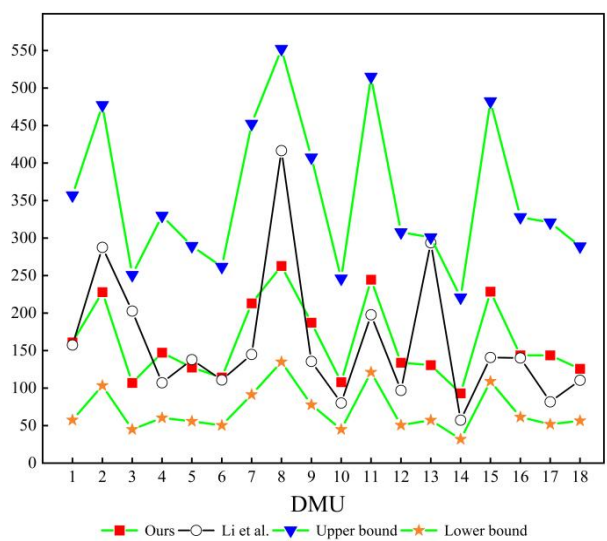

**Figure 3** Upper and lower bounds of revenue and cost allocation

## 5. Conclusions

Most existing studies on the issue of resource allocation and cost-sharing treat resources as additional inputs, and propose different allocation schemes based on the principle of efficiency invariant or efficiency maximization. In this paper, according to the essence and practical significance of resource allocation, we consider the common revenue as an exogenous variable and propose a modified Shapley value method based on the cross-efficiency.

Firstly, we solve the uniqueness problem of cross-efficiency by adopting the DEA cross-efficiency method considering both competition and cooperation among DMUs. Thus, the unique cross-efficiency matrix is obtained, and then the cooperative game model is defined from the cross-efficiency matrix based on the cooperation among the DMUs. Secondly, by fully considering the rationality and fairness of the allocation scheme, we modify the Shapley value of the cooperative game, so that the allocated common revenue of the DMU is positively correlated to the degree of its contribution to the organization. According to the characteristics of the cross-efficiency and the modified Shapley values, combining with the realistic situation, we also define the upper and lower bounds of the Shapley values from the optimistic and pessimistic perspective, respectively. Then, according to the modified Shapley value and its upper and lower bounds, we obtain the allocation result and the corresponding upper and lower bounds for each DMU. Finally, we illustrate the computational process of our proposed allocation scheme through a numerical case. Furthermore, comparing to the literature [25], we apply the method to the empirical case of a city commercial bank with 18 branches, and show that the upper and lower bounds of our allocation scheme are reasonable and valid.

The common revenue allocation method proposed in this paper can be further extended. On the one hand, we can consider the internal revenue allocation problem of multi-stage DMUs, and construct the corresponding revenue allocation model combined with the characteristics of cross-efficiency. On the other hand, this paper allocates the common revenue from the perspective of the contribution of DMUs. The allocation methods can also be proposed based on the degree of satisfaction of DMUs or to establish a comprehensive revenue allocation model considering both contribution and satisfaction, which will be a very meaningful research topic in the future.

**Acknowledgement**
This work was supported by National Natural Science Foundation of China (Grant No. 72001207).